\documentclass[preprintnumbers,showkeys,amsmath,amssymb]{revtex4}
\usepackage{graphicx}
\usepackage{dcolumn}
\usepackage{bm}

\usepackage{epstopdf}
\usepackage{color}

\begin{document}

\preprint{}

\title{Longitudinal relaxation of a nitrogen-vacancy center in a spin bath by generalized cluster-correlation expansion method}

\author{Zhi-Sheng Yang$^{1}$ ,Yan-Xiang Wang$^{1}$ ,Ming-Jie Tao$^{1}$ ,Wen Yang$^{2}$, Mei Zhang$^{1}$ ,Qing Ai$^{1}$\footnote{Corresponding author:
aiqing@bnu.edu.cn} and Fu-Guo Deng$^{1}$ }

\address{$^{1}$Department of Physics, Applied Optics Beijing Area Major Laboratory,
Beijing Normal University, Beijing 100875, China\\
$^{2}$Beijing Computational Science Research Center, Beijing 100094, China}

\begin{abstract}
We theoretically study the longitudinal relaxation of a nitrogen-vacancy (NV) center
surrounded by a $^{13}$C nuclear spin bath in diamond.
By incorporating electron spin in the cluster, we generalize the cluster-correlation expansion (CCE) to theoretically simulate the population dynamics of electron spin of NV center.
By means of the generalized CCE,
we numerically demonstrate the decay process of electronic
state induced by cross relaxation at the ambient temperature. It is shown
that the CCE method is not only capable of describing pure-dephasing
effect at large-detuning regime, but it can also simulate the quantum
dynamics of populations in the nearly-resonant regime.
\end{abstract}


\maketitle

\section{Introduction}
\label{sec:Introduction}

Nitrogen-vacancy (NV) center is a pair of point defects in the diamond lattices \cite{Doherty13,Dobrovitski13}. It is comprised of a substitutional nitrogen atom and a vacancy at the adjacent lattice.
Because of possessing $S=1$ electron spin, the ground-state manifold of the negatively-charged NV center is triplet states
and it can be optically initialized and measured.

For a sample of high purity, the NV center possesses a long coherence time and can be further prolonged to approach one second by dynamical decoupling \cite{deLange10,Bar-Gill13}.
It can be coherently controlled by microwave fields and is highly sensitive to external electric and magnetic fields.
These superior characteristics make the NV center a promising candidate for
quantum information processing \cite{Childress06,Dutt07,Song16-1,Wang18} and quantum sensing of electric and magnetic fields \cite{Dolde11,Maze08,Tetienne13}, temperature \cite{Toyli13}, stress \cite{Grazioso13},
biological structures \cite{Shi15} and chemical reactions \cite{Liu17}. Such appealing applications require full knowledge of its decoherence behavior,
including both transverse and longitudinal relaxation \cite{Hall16,WangP15b,WangH15},
over a broad range of physical parameters,
 such as temperature, magnetic field, type and concentration of impurities. Among quantum many-body theories for many-spin bath dynamics \cite{Ishizaki09,Ai13}, e.g. the density matrix cluster expansion \cite{Witzel06,Witzel07a,Witzel07b}, the pair correlation approximation \cite{Yao06,Yao07,Liu07}, and the linked-cluster expansion \cite{Saikin07},
 the cluster-correlation expansion (CCE) \cite{Yang08,Yang09,Li17}
 developed by one of the authors WY has been successfully applied to
 describe the pure-dephasing process of a central spin
 coupling to a many-spin bath in the largely-detuned regime \cite{Yang15}.
 The anomalous decoherence effect of NV center in a quantum spin bath
 has been theoretically predicted \cite{Zhao11-2} and experimentally verified \cite{Huang11}.
 Interestingly, the decoherence of NV center,
 which was previously considered as an obstacle to realize quantum computation \cite{Jacques09,Song16-2}, has been explored to detect a remote nuclear spin dimer \cite{Zhao11-1} and proposed for navigation by using the weak geomagnetic field \cite{Li17}.

Meanwhile, the experimental investigation on the longitudinal relaxation of NV center has been delicately performed for a broad range of temperature, magnetic field, concentration of impurities, ever since 1991 \cite{Redman91,Jarmola12,Jarmola15,Mrozek15}.
The physical mechanism of longitudinal relaxation above 100K is attributed to Orbach and Raman phonon process \cite{Redman91,Jarmola12}.
When the temperature reaches as low as a few tens of kelvins, $T_1$ clearly demonstrates dependence on the concentration of magnetic impurities \cite{Jarmola12}. In addition, when varying the magnetic field,
$T_1$ abruptly drops around some energy-level anti-crossing points \cite{Jarmola12}. These observations motivate us to develop a theory to
describe the longitudinal relaxation of NV center in a quantum many-spin bath. In this paper, by including the electron spin in the cluster expansion, we generalize the CCE to simulate the quantum dynamics of NV center exchanging population with a nuclear spin bath in the nearly-resonant regime.

When a static magnetic field is applied to make the NV center close to the
level anticrossing point, the hyperfine interaction of the NV center with
the $^{13}$C nuclear spins induces the NV electron spin relaxation. The numerical simulation with the CCE approach yields convergent result and thus is capable of fully describing the NV electron spin relaxation induced by the $^{13}$C spin bath.

This paper is organized as follows: In Sec.~\ref{sec:Model},
we present our system model. In Sec.~\ref{sec:CCE},
we show the CCE approach after generalization.
In Sec.~\ref{sec:LongRelax}, we simulate the longitudinal relaxation
of NV center surrounded by $^{13}$C nuclear bath with the generalized CCE.
A detailed discussion and a summary are enclosed in Sec.~\ref{sec:Discussion}.
\section{Model}
\label{sec:Model}

\begin{figure*}[!ht]
\includegraphics[width=18 cm,angle=0]{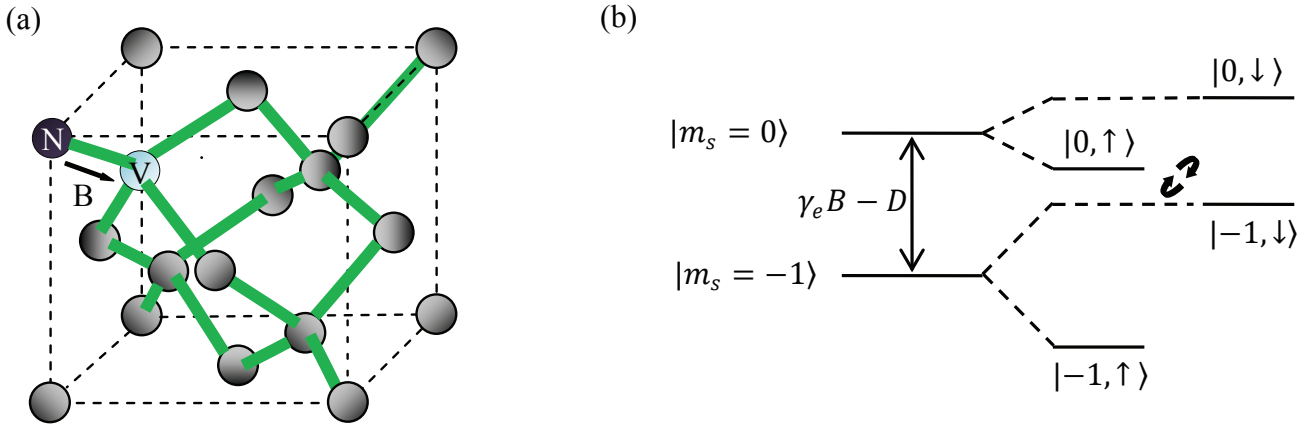}
\caption{(color online)
Schematic illustration of an NV center surrounded by a spin
bath. (a) An NV center in the crystal structure of a diamond, where a static
magnetic field is applied along the principal axis of NV center.
(b) The hyperfine structure of an NV center and one $^{13}$C nuclear
spin in the bath. }
\label{fig:Scheme}
\end{figure*}

As shown in Fig.~\ref{fig:Scheme}, a negatively-charged NV center is coupled
to the $^{13}$C nuclear spins randomly located at the diamond lattices, with
the natural abundance of $^{13}$C being $1.1\%$ \cite{Zhao12}. Here, since
we focus on the longitudinal relaxation induced by $^{13}$C nuclear spins
far away from the NV center, in our numerical simulation we use the
randomly-generated bath in which there is no $^{13}$C nuclear spin in the
vicinity of NV center \cite{Zhao12}. The NV center consists of a
substitutional nitrogen atom adjacent to a vacancy. The ground state of NV
center is a triplet state with $m_{s}=0$ and $m_{s}=\pm1$ denoted by $%
\vert0\rangle$ and $\vert\pm1\rangle$ respectively. When there is no
magnetic field applied to the NV center, $\vert\pm1\rangle$ are degenerate
and they are separated from $\vert0\rangle$ by zero-field splitting $D=2.87$%
GHz \cite{Doherty13,Dobrovitski13}. In an external magnetic field $B_z$
along the NV axis, the degeneracy between $m_{s}=\pm1$ levels are lifted and
thus the full Hamiltonian of the total system describing the NV center and
spin bath is
\begin{eqnarray}
H & = & H_{\mathrm{NV}}+H_{\mathrm{bath}}+H_{\mathrm{int}},  \label{eq:H}
\end{eqnarray}
where the Hamiltonian of the NV center has the form
\begin{eqnarray}
H_{\mathrm{NV}} & = & DS_{z}^{2}-\gamma_{\mathrm{e}}B_{z}S_{z}
\end{eqnarray}
with $\gamma_{\mathrm{e}}=-1.76\times10^{11}$ rad$\cdot$s$^{-1}$T$^{-1}$
being the electronic gyromagnetic ratio \cite{Zhao12} and $%
S_{z}=\vert1\rangle\langle1\vert-\vert-1\rangle\langle-1\vert$.

The Hamiltonian of the bath reads
\begin{eqnarray}
H_{\mathrm{bath}}&=&\sum_{i<j}D_{ij}%
\left[\mathbf{\mathbf{\mathbf{I}}}_{i}\cdot\mathbf{\mathbf{I}}_{j}-\frac{%
3\left(\mathbf{\mathbf{I}}_{i}\cdot\mathbf{r}_{ij}\right)\left(\mathbf{%
\mathbf{r}}_{ij}\cdot\mathbf{\mathbf{I}}_{j}\right)}{r_{ij}^{2}}\right] \nonumber \\
&&-\gamma_{\mathrm{c}}B_{z}\sum_{i}I_{i}^{z},
\end{eqnarray}
where the gyromagnetic ratio of $^{13}$C nuclear spin is \cite{Zhao12} $%
\gamma_{\mathrm{c}}=6.73\times10^{7}$rad$\cdot$s$^{-1}$T$^{-1}$,
\begin{equation}
D_{ij}=\frac{\mu_{0}\gamma_{\mathrm{c}}^{2}}{4\pi r_{ij}^{3}}\left(1-\frac{3%
\mathbf{\mathbf{r}}_{ij}\mathbf{\mathbf{r}}_{ij}}{r_{ij}^{2}}\right)
\end{equation}
is the magnetic dipole-dipole interaction between the $i$th and $j$th $^{13}$%
C nuclear spins with $\mathbf{\mathbf{r}}_{ij}$ being the displacement
vector from $i$th to $j$th spins and $\mu_{0}$ being the vacuum permeability.

The electron spin and nuclear spins are coupled by hyperfine interaction.
Because we are interested in a large number of $^{13}$C nuclear spins far
away from the NV center, the hyperfine interaction is mainly described by
magnetic dipole-dipole interaction and thus the interaction Hamiltonian is
\begin{equation}
H_{\mathrm{int}}=\mathbf{S}\cdot\sum_{i}A_{i}\cdot\mathbf{\mathbf{I}}_{i},
\end{equation}
where
\begin{equation}
A_{i}=\frac{\mu_{0}\gamma_{\mathrm{c}}\gamma_{\mathrm{e}}}{4\pi r_{i}^{3}}%
\left(1-\frac{3\mathbf{\mathbf{r}}_{i}\mathbf{\mathbf{r}}_{i}}{r_{i}^{2}}%
\right)
\end{equation}
is the hyperfine coupling tensor, $\mathbf{r}_i$ is the position vector of $%
i $th nuclear spin, and the position of NV center is chosen as the origin.

\section{Generalized Cluster-Correlation Expansion}
\label{sec:CCE}
In this section, by taking electron spin into consideration in the cluster, we generalize the CCE method \cite{Yang08,Yang09} to describe the electron spin relaxation in a spin bath. Traditional CCE method has been successfully applied to calculating the pure dephasing of an NV center in the large-detuning regime. However, the central spin flip must be considered when the energy relaxation of the NV center is involved in the nearly-resonant regime.

\begin{figure*} [!ht]
\begin{center}
\includegraphics[width=18 cm,angle=0]{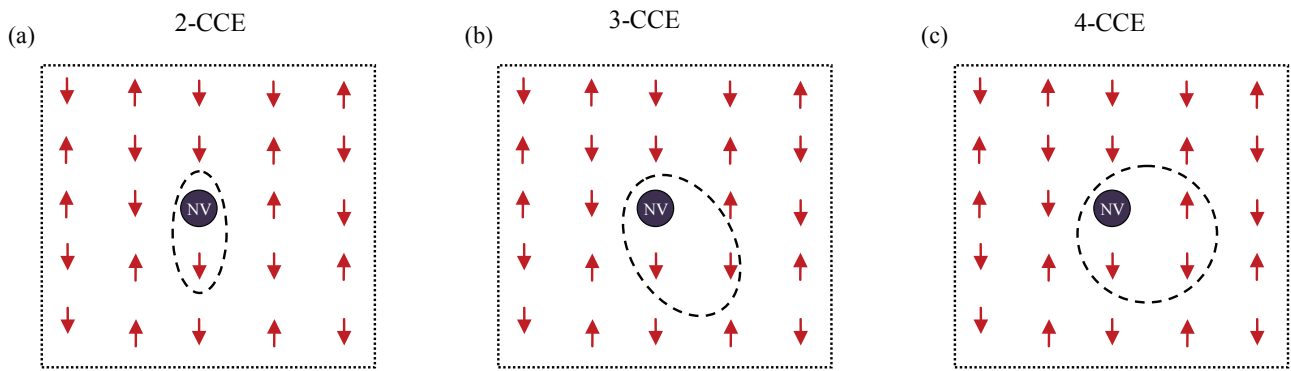}
\caption{(color online)
Schematic of generalized CCE. The bath of $^{13}$C nuclear spins and
 the NV center forms an open quantum system. The $^{13}$C
atoms with $1/2$ nuclear spin denoted by the small red arrows are randomly located on the diamond lattice. (a) 2-CCE: A cluster containing one bath spin and the NV center. (b) 3-CCE: A cluster containing two bath spins and the NV center. (c) 4-CCE: A cluster containing three bath spins and the NV center.
The spins outside the cluster are assumed to be frozen when calculating the cluster contribution.}
\label{fig:CCE-bath}
\end{center}
\end{figure*}

Assuming that the bath consists of spin $i$ only, cf.  Fig.~\ref{fig:CCE-bath}(a), the probability $P(t)$ of NV center in the state $|0\rangle$ is explicitly calculated as
\begin{equation}
\tilde{P}_{\{i\}}=P_{\{i\}}\equiv \frac{\text{Tr}\left(\rho
(0)e^{iH_{\{i\}}t}|0\rangle \langle 0|e^{-iH_{\{i\}}t}\right)}{\text{Tr}\left(\rho
(0)e^{iH_{\mathrm{NV}}t}|0\rangle \langle 0|e^{-iH_{\mathrm{NV}}t}\right)},
\label{eq:1CCE}
\end{equation}%
where $\rho(0)$ is the initial state of the total system including the NV center and nuclear spin bath,
\begin{equation}
H_{\{i\}}=H_{\mathrm{NV}}-\gamma _{\mathrm{c}}B_{z}I_{i}^{z}+\mathbf{S}\cdot
A_{i}\cdot \mathbf{\mathbf{I}}_{i}
\end{equation}%
is obtained from Eq.~(\ref{eq:H}) by dropping all terms other than spin $i$.

Assuming that the bath consists of spin $i$ and spin $j$, cf.  Fig.~\ref{fig:CCE-bath}(b), $P(t)$ reads
\begin{equation}
P_{\{i,j\}}\equiv\frac{\text{Tr}\left(\rho(0)
e^{iH_{\{i,j\}}t}\vert0\rangle\langle0\vert e^{-iH_{\{i,j\}}t}\right)} {\text{Tr}%
\left(\rho(0) e^{iH_{\mathrm{NV}}t}\vert0\rangle\langle0\vert e^{-iH_{\mathrm{NV}%
}t}\right)},  \label{eq:2CCE}
\end{equation}
where
\begin{eqnarray}
H_{\{i,j\}}&=& H_{\mathrm{NV}}-\gamma_{\mathrm{c}}B_{z}\sum_{\alpha=i,j}I_{\alpha}^{z}%
+\mathbf{S}\cdot \sum_{\alpha=i,j}A_{\alpha}\cdot\mathbf{\mathbf{I}}_{\alpha} \nonumber \\
&&+D_{ij}\left[\mathbf{\mathbf{\mathbf{I}}}_{i}\cdot\mathbf{%
\mathbf{I}}_{j}-\frac{3\left(\mathbf{\mathbf{I}}_{i}\cdot\mathbf{r}%
_{ij}\right)\left(\mathbf{\mathbf{r}}_{ij}\cdot\mathbf{\mathbf{I}}_{j}\right)}{r_{ij}^{2}}\right]
\end{eqnarray}
is calculated from Eq.~(\ref{eq:H}) by dropping all terms other than spin $i$
and spin $j$, and the spin-pair correlation is
\begin{equation}
\tilde{P}_{\{i,j\}}\equiv \frac{P_{\{i,j\}}}{\tilde{P}_{\{i\}}\tilde{P}%
_{\{j\}}}.
\end{equation}

Assuming that the bath consists of three spins $i$ and $j$ and $k$, cf.  Fig.~\ref{fig:CCE-bath}(c), $P(t)$
in this case becomes
\begin{equation}
P_{\{i,j,k\}}\equiv\frac{\text{Tr}\left(\rho(0)
e^{iH_{\{i,j,k\}}t}\vert0\rangle\langle0\vert e^{-iH_{\{i,j,k\}}t}\right)} {\text{%
Tr}\left(\rho(0) e^{iH_{\mathrm{NV}}t}\vert0\rangle\langle0\vert e^{-iH_{\mathrm{%
NV}}t}\right)},  \label{eq:3CCE}
\end{equation}
where
\begin{align}
&H_{\{i,j,k\}}=H_{\mathrm{NV}}-\gamma_{\mathrm{c}}B_{z}\sum_{\alpha=i,j,k}I_{%
\alpha}^{z} +\mathbf{S}\cdot\sum_{\alpha=i,j,k}A_{\alpha}\cdot\mathbf{\mathbf{I}}_{\alpha} \nonumber \\
&+\sum_{\alpha=i,j,k}\sum_{\beta>\alpha}D_{\alpha\beta}\left[%
\mathbf{\mathbf{\mathbf{I}}}_{\alpha}\cdot\mathbf{\mathbf{I}}_{\beta} %
-\frac{3\left(\mathbf{\mathbf{I}}_{\alpha}\cdot\mathbf{r}_{\alpha\beta}\right)\left(%
\mathbf{\mathbf{r}}_{\alpha\beta}\cdot\mathbf{\mathbf{I}}_{\beta}\right)}{%
r_{\alpha\beta}^{2}}\right]
\end{align}
is calculated from Eq.~(\ref{eq:H}) by dropping all terms other than the
three spins $i$ and $j$ and $k$, and the three-spin correlation is
\begin{equation}
\tilde{P}_{\{i,j,k\}}\equiv \frac{P_{\{i,j,k\}}}{\tilde{P}_{\{i\}}\tilde{P}%
_{\{j\}}\tilde{P}_{\{k\}}\tilde{P}_{\{i,j\}}\tilde{P}_{\{j,k\}}\tilde{P}%
_{\{k,i\}}}.
\end{equation}

For the bath with arbitrary number of spins, $P(t)$ is generalized as
\begin{equation}
P_{\mathfrak{c}}\equiv\frac{\text{Tr}\left(\rho(0) e^{iH_{\mathfrak{c}%
}t}\vert0\rangle\langle0\vert e^{-iH_{\mathfrak{c}}t}\right)} {\text{Tr}\left(\rho(0)
e^{iH_{\mathrm{NV}}t}\vert0\rangle\langle0\vert e^{-iH_{\mathrm{NV}}t}\right)},
\label{eq:MCCE}
\end{equation}
where $H_{\mathfrak{c}}$ is obtained from Eq.~(\ref{eq:H}) by dropping all
terms other than spins in the cluster $\mathfrak{c}$, and the spin-cluster
correlation is
\begin{equation}
\tilde{P}_{\mathfrak{c}}\equiv \frac{P_{\mathfrak{c}}}{\prod_{\mathfrak{c}%
^\prime\subset\mathfrak{c}}\tilde{P}_{\mathfrak{c}^\prime}}.
\end{equation}
Furthermore, in Eqs.~(\ref{eq:1CCE},\ref{eq:2CCE},\ref{eq:3CCE},\ref{eq:MCCE}),
all of the denominators are equal to unity because the system is
initially prepared at $\vert0\rangle$, which is also the eigenstate of $H_{\mathrm{NV}}$.

Generally speaking, it is impossible to exactly calculate $P(t)$ for a large
number of spins as the dimension of Hilbert space scales exponentially with
the number of spins. The $M$-CCE method approximates $P(t)$ as
\begin{equation}
P^{(M)}=\prod_{\vert\mathfrak{c}\vert\leq M}\tilde{P}_{\mathfrak{c}},
\end{equation}
where $\vert\mathfrak{c}\vert$ is the number of spins in the cluster $\mathfrak{c}$. For example, the first-order truncation of $P(t)$ yields
\begin{equation}
P^{(1)}=\prod_i\tilde{P}_{\{i\}}.
\end{equation}
And the second-order truncation of $P(t)$ reads
\begin{equation}
P^{(2)}=\prod_i\tilde{P}_{\{i\}}\prod_{i,j}\tilde{P}_{\{i,j\}}.
\end{equation}

\section{Decay of NV Center Induced by Nuclear Spin Bath}
\label{sec:LongRelax}
In this section, we discuss the longitudinal relaxation of electron
spin of NV center induced by coupling to the nuclear spin bath.
When the energy gap between $\vert0\rangle$ and $\vert-1\rangle$ approaches the energy gap of $^{13}$C nuclear spins,
e.g. by tuning the magnetic field, the electron spin
exchanges polarization with nuclear spins, cf. Fig.~\ref{fig:Scheme}(c), and
the longitudinal relaxation of electronic spin occurs. For simplicity, we consider the situation of low temperature and a single NV in the diamond, so that the decay of electron spin is mainly due to hyperfine interaction
with nuclear spins, and the effect of diamond lattice phonons on the
relaxation is beyond the scope of the present investigation.

The state of electronic spin can be optically initialized into
$|0\rangle $, corresponding to the density matrix
\begin{equation}
\rho _{\mathrm{NV}}=|0\rangle \langle 0|.
\end{equation}%
In principle, the $^{13}$C nuclear spin bath should be in the
thermal equilibrium state since the typical experimental
temperature is much higher than the nuclear spin Zeeman splitting, even in a
strong magnetic field, e.g. several T. Therefore, without loss of generality, we take the density matrix of the $^{13}$C bath as
\begin{equation}
\rho _{\mathrm{B}}=\frac{1}{2^N}\prod_{i=1}^{\otimes N}\left(|\uparrow \rangle _{i}\langle\uparrow |
+|\downarrow \rangle _{i}\langle\downarrow |\right),
\end{equation}%
where $|\uparrow \rangle _{i}$ ($|\downarrow \rangle _{i}$) is the spin-up
(spin-down) state of the $i$th $^{13}$C nuclear spin, with the quantization
axis along the N-V symmetry axis. As a result, the density matrix of total
system at the initial time is
\begin{equation}
\rho \left( 0\right) =\rho _{\mathrm{NV}}\otimes \rho _{\mathrm{B}}.
\end{equation}%
The evolution of the coupled system is given by
\begin{equation}
\rho \left( t\right) =e^{-iHt}\rho \left( 0\right) e^{iHt}.
\end{equation}%
By partially tracing over the degrees of freedom of the bath, we could
obtain the survival probability of the initial state $|0\rangle $ of the NV
electron spin as
\begin{equation}
P(t)=\mathrm{Tr}_{\mathrm{B}}\langle 0|e^{-iHt}\rho \left( 0\right) e^{iHt}|0\rangle .
\label{eq:P}
\end{equation}

\begin{figure*} [!ht]
\centering
\includegraphics[width=18 cm,angle=0]{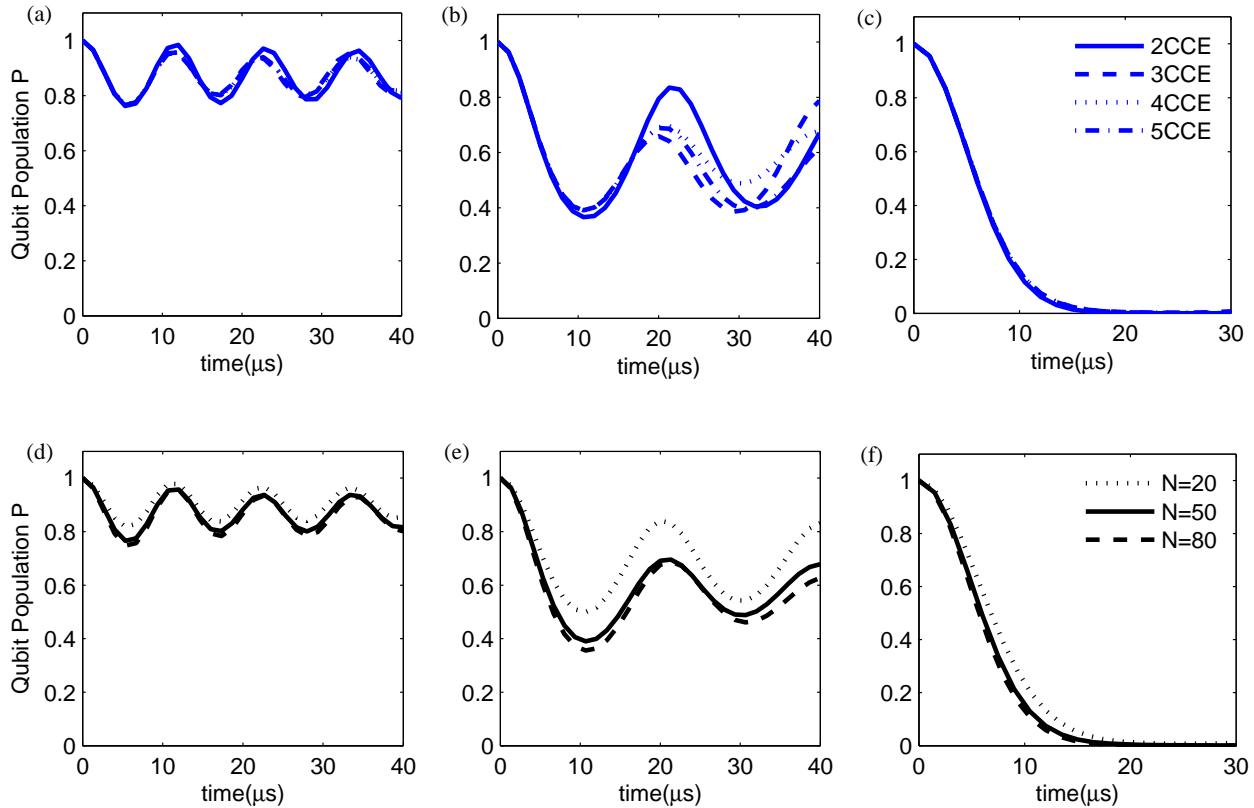}
\caption{(color online) The longitudinal relaxation process of the NV center under different magnetic fields.
Survival probability of initial state of electron spin by different orders of CCE with $N=50$ and (a) $B_z=1025.01$G, (b) $B_z=1024.99$G,
(c) $B_z=1024.97$G.
Survival probability with different bath sizes and (d) $B_z=1025.01$G, (e) $B_z=1024.99$G, (f) $B_z=1024.97$G.}
\label{fig:relaxation}
\end{figure*}

For a small spin bath, we can exactly calculate the longitudinal relaxation
of electron spin via Eq.~(\ref{eq:P}). However, with increasing size of the
nuclear spin bath, this approach quickly becomes unfeasible, because the
dimension of the Hilbert space grows exponentially with the number of nuclei
\cite{Dobrovitski09,Zhang07}. In this case, an approximate many-body theory
with high performance has to be considered. In previous works, the CCE
theory has been successfully applied to describe the pure dephasing of a
central spin in a spin bath \cite{Yang08,Yang09,Zhao11-1,Zhao11-2}. Here, we
generalize the CCE theory to deal with the longitudinal relaxation of the
electron spin in a spin bath. In Fig.~\ref{fig:relaxation}(a),
we numerically simulate the survival probability of the initial state of
electron spin surrounded by $N=50$ nuclear spins under the magnetic
field intensity $B_z=1025.01454$G along the NV axis. When the order of CCE
method is increased, the results quickly converge, e.g., the results from
4th-order CCE and 5th-order CCE are indistinguishable. This suggests that
4th-order CCE already offers a reliable solution to the quantum dynamics of
central spin under the influence of a many-spin bath. Figure~\ref{fig:relaxation}(a) shows the population of electron spin experiences a damped oscillation with a small amplitude
because there is still a large energy gap between the electron spin and nuclear spin as compared to their hyperfine interaction.
When we reduce the energy gap of electron spin by tuning the magnetic field, cf. Fig.~\ref{fig:relaxation}(b), both the amplitude and period of damped oscillation becomes larger as the electron and nuclear spins are getting close to resonance. If we further tune the electronic level to the nearly-resonant case, as shown in Fig.~\ref{fig:relaxation}(c), the population of electronic initial state decays to zero around 10$\mu$s without collapse and revival. When the electron spin resonantly interacts with a nuclear spin, its population will demonstrate Rabi oscillation with full amplitude and frequency $A_i$. Since all nuclear spins interact with the electron spin with different hyperfine interactions,
the electronic population will irreversibly decay and no revival will occur.
Furthermore, we explore the effect of the bath's size on the relaxation.
In Fig.~\ref{fig:relaxation}(d), the population dynamics of the electron spin is investigated for $^{13}$C baths of different sizes with a large energy gap. When the size of the bath is increased, the difference before and after the change becomes more and more negligible. As further enlarging the size will not significantly decrease the difference but result in much more computation time, the 4-CCE theory with a bath of $N=50$ nuclear spins already yields a reliable result, as confirmed in Fig.~\ref{fig:relaxation}(e) and (f).

\begin{figure*}[!ht]
\centering
\includegraphics[width=8 cm,angle=0]{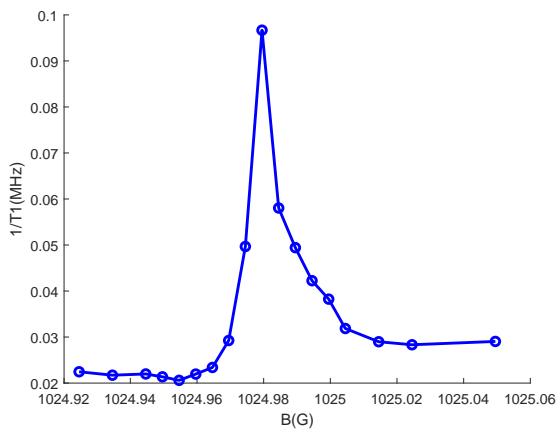}
\caption{(color online) The longitudinal electron-spin relaxation time ($T_\mathrm{1}$) of the NV center under different magnetic fields at 4CCE.}
\label{fig:Time1}
\end{figure*}

Furthermore, we numerically fit the relaxation process by an exponential decay, and show the longitudinal electron-spin relaxation time ($T_\mathrm{1}$) under different magnetic fields at 4CCE and $N=50$ in Fig.~\ref{fig:Time1}. It shows that $1/T_\mathrm{1}$ varies sensitively in response to the change of the magnetic field. Especially, when the magnetic field is tuned to the resonance point, i.e. 1024.975~G, $1/T_\mathrm{1}$ has been increased by nearly one order of magnitude. This result is reasonable since around the resonance, many different nuclear spins can exchange spin with the electron with different frequencies and thus result in enhanced decoherence. It is consistent with experimental investigation shown in Ref.~\cite{Jarmola12}, where electronic decoherence around the resonance was enhanced by coupling to electron-spin bath. 

\section{DISCUSSION AND SUMMARY}
\label{sec:Discussion}

In this paper, by embodying the electron spin in the cluster,
we generalize the CCE approach to deal with the longitudinal
relaxation of a central electron spin due to its couplings to a
nuclear spin bath in diamond. In the largely-detuned case,
because the electron spin can not effectively exchange energy with
the nuclear spins, the couplings to the nuclear spin bath manifest
as the pure-dephasing phenomenon of the NV center \cite{Zhao12}.
Since the pure-dephasing effect is sensitive to the local magnetic field
at the NV center, it can be fully utilized to detect magnetic source
with internal structure \cite{Zhao11-1}.

When getting close to the degenerate point, we show diverse decay phenomena of initial state's population by tuning the magnetic field.
When the energy gap of electronic levels are small but not too small, only a few nuclear spins can exchange population with the electron spin.
It results in a damped oscillation with a very small decay rate.
If we further reduce the energy gap, a more remarkable damped oscillation be observed as more nuclear spins can be probed by the electron spin.
When the electron spin is finally tuned to be in resonance with all nuclear spins, the population of initial state will monotonically decay and no revival will take place.

We consider the feasibility with the current accessible parameters in the
experiments. The state of NV center $|0\rangle$ can be initialized by applying 532 nm laser \cite{Doherty13}. In order to bring the electronic spin close to the degenerate point, a magnetic field with strength $B_z\approx 1025$ G should be applied to NV center, which is available at current laboratory conditions \cite{Lillie17}. After the optical pumping, the NV center will be left to evolve under the influence of nuclear spin bath for an interval and then readout by detecting the fluorescence after spin-dependent optical excitation \cite{Jarmola04}.


In this paper, because of the generalization,
the CCE approach is capable of describing the quantum evolution of
the whole density matrix of the system including both the off-diagonal and diagonal terms.
Recently, quantum coherent energy transfer has attracted broad interest from different disciplines.

\section*{ACKNOWLEDGMENTS}

We thank L. P. Yang for helpful discussions. F. G. Deng was supported by the
National Natural Science Foundation of China under Grant No.~11474026 and
the Fundamental Research Funds for the Central Universities under Grant
No.~2015KJJCA01. W. Yang was supported by the
National Natural Science Foundation of China under Grant No.~11274036 and No.~11322542, and the MOST under Grant No.~2014CB848700.
M. Zhang was supported by the National Natural Science Foundation of China under Grant No.~11475021. Q. Ai was supported by the National Natural Science Foundation of China under Grant No.~11505007, and the Open Research Fund Program of the State Key Laboratory of Low-Dimensional Quantum Physics, Tsinghua University under Grant No.~KF201502.

\end{document}